\documentclass[11pt]{snmp2003a}

\usepackage{latexsym,amsmath,amssymb,amsthm}

\newcommand{\C}{{\mathbb C}}
\newcommand{\Cl}{{\mathbb{C}\ell}}
\newcommand{\End}{{\rm End}\ }
\newcommand{\g}{{\gamma}}
\newcommand{\G}{{\Gamma}}

\newcommand{\R}{{\mathbb R}}

\newcommand{\s}{{\sigma}}

\newcommand{\one}{{1\kern-2.5pt \text{l}} }
\newcommand{\cg}{{\cal G}}

\newcommand{\Spin}{{\rm Spin}}

\begin{document}

%\FirstPageHeading{Budinich}
% The parameter is the label of the article. Good choice is the last name of
%the first author

%\ShortArticleName{Division Algebras in Pure Spinor Geometry}
% maximum 75 symbols

\ArticleName{Internal Symmetry From Division Algebras in Pure
Spinor Geometry}

% Names of the authors for the title of the paper
\Author{P. Budinich~$^\dag$} %\AuthorNameForHeading{P. Budinich}
%\AuthorNameForContents{BUDINICH P.}
%\ArticleNameForContents{Internal Symmetry From Division Algebras
%in Pure Spinor Geometry}

% Address of First Author
\Address{$^\dag$~International School for Advanced Studies,
Trieste, Italy} \EmailD{fit@ictp.trieste.it}
% Address of Second Author
%\Address{$^\ddag$~Address of Second Author, Country}
%\EmailD{email@address}

% In the case of the same organization, please use the following standard
%\Author{First Names LASTNAME and Second COAUTHOR}
%\AuthoqNameForHeading{F.N. Lastname and S. Coauthor}
%\AuthorNameForContents{LASTNAME F.N. and COAUTHOR S.}
%\ArticleNameForContents{Example Article for the Proceedings of the Fifth
%International Conference  ``Symmetry  in Nonlinear Mathematical Physics''}
%\Address{Address of Author(s), Country}
%\Email{email1@address, email2@address}

\Abstract{The E'. Cartan's equations defining ``simple'' spinors
(renamed ``pure'' by C. Chevalley) are interpreted as equations of
motions for fermion multiplets in momentum spaces which, in a
constructive approach based bilinearly on those spinors, result
compact and lorentzian, naturally ending up with a ten dimension
space.

The equations found are most of those traditionally adopted ad hoc
by theoretical physics in order to represent the observed
phenomenology of elementary particles. In particular it is shown
how, the known internal symmetry groups, in particular those of
the standard model, might derive from the 3 complex division
algebras correlated with the associated Clifford algebras. They
also explain the origin of charges, the tendency of fermions to
appear in charged-neutral doublets, as well as the origin of
families.

The adoption of the Cartan's conjecture on the non elementary
nature of euclidean geometry (bilinearly generated by simple or
pure spinors) might throw light on several problematic aspects
of particle physics.}

\section{Foreword}

It is well known that fermions and bosons are to be conceived as
the quanta of spinor- and of euclidean tensor-fields,
respectively. Now it is generally agreed that bosons may be always
bilinearly represented in terms of fermions which then appear as
the elementary constituents of matter. The consequent physical
assumption is that bosons should be bound states of  fermions,
having in mind the example of nuclei, say, bound states of
nucleons.

However this interpretation appears some times not viable; as in
the case of the photon which, notoriously, may not be conceived as
a bound state of neutrinos as proposed a long time ago by L. de
Broglie. Then one may shift back the attention to the classical
fields, before quantization, and try to see if possibly the
euclidean tensor fields might be steadily bilinearly represented
in terms of spinor fields, and, consequently, the physical,
somehow naive, explanation in terms of bound states might be
substituted by the purely geometrical one which, by itself, may
offer aspects of great interest and of deep meanings, since it
could reveal, among others, a striking parallelism between
geometry and physics, as we will see. This way was attempted by
several distinguished authors, starting from W. Heisenberg.

In this paper we will try to show that indeed the geometrical way
might be the right one, provided the spinors adopted are those
which \'E. Cartan named simple~\cite{Budinich:Cartan37}, later
renamed pure by E. Chevalley~\cite{Budinich:Chevalley54}. In fact
\'E. Cartan himself advocated the conjecture that euclidean
geometry might have to be conceived as not elementary insofar its
elementary constituents might be represented by simple spinor.

In the following we will assume the reader familiar with spinor
geometry rich of published literature~\cite{Budinich:Cartan37},
\cite{Budinich:Chevalley54},
\cite{Budinich:Budinich&Trautman1989},
\cite{Budinich:Penrose&Rindler84} and try to concentrate on the
synthetic exposition of some results.

\section{Basic definitions and propositions}

Given a space $W=\C^{2n}$ with Clifford algebra $\C(2n)=\End S$
with generators $\gamma_a$ obeying:
\begin{equation*}
[\g_a,\g_b]_+=2\delta_{ab},\qquad a,b,=1,2,\dots 2n
\end{equation*}
let $\psi_D\in S$  represent a $2^n$ dimensional (Dirac) spinor.
For $z \in W$, the Cartan's equation:
\begin{equation}
z_a\g^a\psi_D=0,\qquad a=1,2,\dots 2n \label{Budinich:equation1}
\end{equation}
defines a totally null plane of dimension $d$, indicated in the
following with $T_d(\psi_D)$.

The volume element $\g_{2n+1}$ defined by:
$\g_{2n+1}=\g_1\g_2\dots\g_{2n}$ anticommutes with all $\g_a$ and
$\g_1,\g_2,\dots\g_{2n},\g_{2n+1}$ generate $\Cl(2n+1)$ whose
associated Pauli spinors will be indicated with $\psi_P$. We will
further indicate with $\varphi_W$ the Weyl spinors defined by:
\begin{equation}
\varphi^\pm_W=\frac{1}{2} (1\pm\g_{2n+1})\psi_D;
\label{Budinich:equation2}
\end{equation}
they are $2^{n-1}$ dimensional and associated with the even
subalgebra $\Cl_0(2n)$ of $\Cl(2n)$. The corresponding Cartan-Weyl
equations will be:
\begin{equation}
z_a\g^a(1\pm\g_{2n+1})\psi_D=0,\qquad a=1,2,\dots 2n
\label{Budinich:equation3}
\end{equation}

\noindent{\bf Definition:} A Weyl spinor $\varphi_W$ associated
with $\Cl_0(2n)$ is simple or pure if the dimension $d$ of the
associated totally null plane: $T_d (\varphi_W )$ is maximal; that
is equal $n$.

\'E. Cartan proved that a simple spinor $\varphi_W$ is equivalent
to $T_n$ ($\varphi_W$) (up to a sign) and he stressed the
importance of this equivalence; in so far it establishes the
fundamental link between spinor-geometry and a specially elegant
and simple sector of euclidean geometry: the projective one; from
which presumably derives the qualification ``simple'' for the
corresponding spinors; now however substituted in the literature
with the word ``pure'' later introduced by Chevalley.

Observe that while the dimension of $T_n (\varphi_W )$ increases
linearly with $n$, those of $\varphi_W$ increase with $n$ as
$2^{n-1}$; therefore for high $n$, simple $\varphi_W$ will have to
be subject to constraint relations; and in fact all Weyl spinors
are simple for $n=1,2,3$ while for $n=4,5,6,7$ simple spinors are
subject to $1,10,66,364$ constraint relations respectively.

We will represent spinors with one column matrices, and $\g_a$
with square ones, and then let us define the main automorphism $B$
of $\Cl(2n)$: $B\g_a=\g^t_aB$; $B\psi =\psi^tB$ where $\g^t_a$ and
$\psi^t$ means $\g_a$ and $\psi$ transposed.

For $\psi ,\phi\in S$ we have~\cite{Budinich:Budinich&Trautman89}:
\begin{equation}
\psi\otimes B\phi =\sum^n_{j=0} T_j=\sum^n_{j=0}
{}_[{}\raisebox{0.5ex}{$\g_{a_1}\g_{a_2}\cdots\g_{a_j}$}{}_]
T^{a_1 a_2\ldots  a_j} ,
\label{Budinich:equation4}
\end{equation}
where the products of $\g_a$ matrices are antisymmetrized and
where:
\begin{equation}
T_{a_1a_2\dots a_j}=\frac{1}{2^n}\langle
B\phi,{}_[{}\raisebox{0.5ex}{$\g_{a_1} \g_{a_2} \cdots
\g_{a_j}$}{}_] \psi\rangle\ .
\label{Budinich:equation5}
\end{equation}

Set in (\ref{Budinich:equation4}) $\phi =\psi =\varphi_W$ and we have:

\begin{proposition} $\varphi_W$ is simple iff in eq.(\ref{Budinich:equation4}):
\begin{equation}
T_0=T_1=\cdots T_{n-1}=0, \quad {\rm while}
\label{Budinich:equation6}
\end{equation}
\begin{equation}
T_n=\frac{1}{2^n} \langle
B\varphi_W,{}_[{}\raisebox{0.5ex}{$\g_{a_1} \g_{a_2} \cdots
\g_{a_n}$}{}_] \varphi_W\rangle\not= 0 .
\label{Budinich:equation7}
\end{equation}
\end{proposition}

Eqs.(\ref{Budinich:equation6}) represent the constraint relations while $T_n$
represents the maximal t.n. plane bilinear in $\varphi_W$ and
equivalent, up to a sign to $\varphi_W$ simple.

Let us now multiply eq.(\ref{Budinich:equation4}) on the left by $\g_a$ and
apply it to
$\g_a\psi$; if we sum over $a$ and set it to zero we obtain:
\begin{equation}
\gamma_a\psi\otimes B\phi\ \gamma^a\psi = z_a\gamma^a\psi =0\quad
a=1,2,\dots 2n
\label{Budinich:equation8}
\end{equation}
that is eq.(\ref{Budinich:equation1}) where:
\begin{equation}
z_a=\frac{1}{2^n} \langle B\phi ,\gamma_a\psi\rangle .
\label{Budinich:equation9}
\end{equation}

We have now the fundamental:

\begin{proposition} Let $z\in W=\C^{2n}$ with components
$z_a=\langle B\phi ,\g_a\psi\rangle$; for $\phi$ arbitrary,
$z_az^a=0$ iff $\psi :=\varphi_W$ is simple or pure.
\end{proposition}

The formal proof is in~\cite{Budinich:Budinich&Trautman89};
however there is also a geometrically visible one. In fact it is
obvious that $z$, defined by (\ref{Budinich:equation9}),
represents the intersection of the planes $T_d ( \phi )$ and
$T_{d'}(\varphi_W )$ now for $\varphi_W$ simple $d'=n$ is maximal
and then $z$ must be null, viceversa if this has to be true for
every $\phi$, it has to be $d'=n$, maximal, and $\varphi_W$ is
simple.

Let us now consider the isomorphism of Clifford algebras:
\begin{equation}
\Cl(2n)\simeq \Cl_0(2n+1)
\label{Budinich:equation10}
\end{equation}
both central simple, and:
\begin{equation}
\Cl(2n+1) \simeq \Cl_0(2n+2)
\label{Budinich:equation11}
\end{equation}
both non simple, from which we have the isomorphisms and
subsequent embeddings of Clifford algebras:
\begin{equation}
\Cl(2n)\simeq \Cl_0(2n+1)\hookrightarrow
\Cl(2n+1)\simeq\Cl_0(2n+2) \hookrightarrow\Cl(2n+2)
\label{Budinich:equation12}
\end{equation}
and the corresponding ones for the associated spinors:
\begin{equation}
\psi_D\simeq\psi_P\hookrightarrow\psi_P\oplus\psi_P\simeq
\psi_W\oplus\psi_W=\Psi_D\simeq\psi_D\oplus\psi_D
\label{Budinich:equation13}
\end{equation}
which implies that a Dirac or Pauli spinor is isomorphic to a
doublet of Dirac or Pauli or Weyl spinors. These isomorphisms may
be represented explicitly. In fact let $\g_a$  be the generators of
$\Cl(2n)$ and $\G_A$ with $A=1,2\dots 2n+2$, those of $\Cl(2n+2)$.
Then for
\begin{equation}
\G^{(m)}_a=\s_m\otimes\g_a\quad m=0,1,2,3; \quad a=1,2,\dots 2n
\label{Budinich:equation14}
\end{equation}
where $\s_0 =1$ and  $\s_1,\s_2,\s_3$  are Pauli matrices, the
corresponding spinor $\psi$  associated with $\Cl(2n+2)$ is:
\begin{equation}
\Psi^{(m)} =\begin{pmatrix} \psi^{(m)}_1\\
\psi^{(m)}_2\end{pmatrix}
\label{Budinich:equation15}
\end{equation}
where, for $m=0$; $1,2$, and $3$ $\psi^{(m)}_1$ and $\psi^{(m)}_2$
are Dirac, Weyl and Pauli spinors respectively. Now defining the
unitary operators:
\begin{equation}
U_m=1\otimes L+\s_m\otimes R=U^{-1}_m,\quad m=0,1,2,3
\label{Budinich:equation16}
\end{equation}
where
\begin{equation}
L=\frac{1}{2} (1+\g_{2n+1});\quad R=\frac{1}{2} (1-\g_{2n+1})
\label{Budinich:equation17}
\end{equation}
we have:
\begin{equation}
U_j\G^{(0)}_AU^{-1}_j=\G^{(j)}_A;\quad U_j\Psi^{(0)}=\Psi^{(j)} ,
\label{Budinich:equation18}
\end{equation}
as easily verified~\cite{Budinich:Budinich02}, from which
\begin{proposition} Dirac and Pauli spinors may be
isomorphically represented by Dirac, Pauli or Weyl spinor doublets.
\end{proposition}

Propositions 2 and 3 represent the basic geometrical tools for our
job. All we have to do is to restrict the above to real spaces,
which will result unambiguously lorentzian, and then to read the
Cartan-Weyl equations we obtain in physical terms. We will see
that indeed the euclidean tensor fields will be bilinear in spinor
fields as already anticipated by Proposition 2; and then their
quanta: the bosons, bilinear in fermions, even in the case they
can not be bound states. In fact we will obtain also Maxwell's
equations from neutrino Weyl equations. Obviously the geometrical
way derives from the basic properties euclidean spaces as
generated by simple spinors in the frame of Cartan's conjecture.
Physical fermions may well be represented by non simple spinors,
for which the bound state approach will be relevant, as in the
case of nuclei.

We will start from the simplest and transparent case of $n=1$ and
then find out the rule for going from $n$ to $n+1$.

\section{The elementary case of n=1. The signature}

Start from $\Cl(2)$ and let $\varphi =\begin{pmatrix} \varphi_0\\
\varphi_1\end{pmatrix}$ and $\psi =\begin{pmatrix} \psi_0\\
\psi_1\end{pmatrix}$ represent two of its associated Dirac
spinors, or Pauli spinors of the isomorphic $\Cl_0(3)$, generated
by the Pauli matrices $\s_1,\s_2\s_3$. Insert them in
eq.(\ref{Budinich:equation4}), where now
$B=-i\s_2=\begin{pmatrix} 0 &-1\\
1& 0\end{pmatrix} :=\epsilon$, which becomes:
\begin{equation}
\begin{pmatrix}
\varphi_0\psi_1&-\varphi_0\psi_0\cr
\varphi_1\psi_1&-\varphi_1\psi_0\end{pmatrix}\equiv \varphi\otimes
B\psi =z_0+z_j\s^j\equiv\begin{pmatrix} z_0+z_3&z_1-iz_2\cr
z_1+iz_2&z_0-z_3\end{pmatrix}
\label{Budinich:equation19}
\end{equation}
from which we easily get both the $z$-vector components bilinear
in the spinors $\psi$ and $\varphi
:z_\mu=\frac{1}{2}\psi^t\epsilon\sigma_\mu\varphi$ (compare the
matrices) and the nullness of the vector $z: z_\mu
z^\mu=z^2_0-z^2_1-z^2_2-z^2_3\equiv 0$ (compute the determinants
of the matrices) in agreement with Proposition 2.

In order to restrict to the real $z_0$ and $z_j$, of interest for
physics, we need to introduce the conjugation operator $C$ defined
by: $C\gamma_a=\bar\gamma_aC,C\varphi =\bar\varphi C$ where
$\bar\gamma_a$ and $\bar\varphi$ mean $\gamma_a$ and $\varphi$
complex conjugate. Then eq.(\ref{Budinich:equation19}) may be
expressed, and uniquely,~\cite{Budinich:Budinich02} in the form:
$$\begin{pmatrix}
\varphi_0\bar\varphi_0&\varphi_0\bar\varphi_1\cr
\varphi_1\bar\varphi_0&\varphi_1\bar\varphi_1\end{pmatrix} =
p_0+p_j\sigma^j =\begin{pmatrix} p_0+p_3&p_1-ip_2\cr
p_1+ip_2&p_0-p_3\end{pmatrix}\hspace{4.5cm} \eqno(19')
$$
and now
\begin{equation}
p_\mu =\varphi^\dagger\sigma_\mu\varphi ,\quad \mu = 0,1,2,3
\label{Budinich:equation20}
\end{equation}
where $\varphi^\dagger$  means  $\varphi$ hermitian conjugate.
Then we have, again identically:
\begin{equation}
p_\mu p^\mu = p^2_0-p^2_1-p^2_2-p^2_3 \equiv 0
\label{Budinich:equation21}
\end{equation}
which shows how $p_\mu$  are the components of a null or optical
vector of a momentum space with Minkowski signature. This is a
particular case of application of Proposition 2. In fact imbed
$\Cl_0(3)$ in the non simple $\Cl(3)$ isomorphic to $\Cl_0(1,3)$
with generators $\gamma_\mu =\left\{ \sigma_1\otimes 1, \ \
-i\sigma_2\otimes\sigma_j\right\}$ and
$\gamma_5=-i\gamma_0\gamma_1\gamma_2\gamma_3=\sigma_3\otimes 1$.
Then we may identify the above Pauli spinor with one of the two
Weyl spinor defined by
\begin{equation}
\varphi_\pm =\frac{1}{2}\left( 1\pm \gamma_5\right)\psi
\label{Budinich:equation22}
\end{equation}
where $\psi$ is a Dirac spinor associated  with $\Cl(1,3)$. Then
eq.(\ref{Budinich:equation20}) identifies with one of the two:
\begin{equation}
p^{(\pm )}_\mu = \tilde\psi\gamma_\mu (1\pm\gamma_5)\psi ;\quad
\mu =0,1,2,3
\label{Budinich:equation23}
\end{equation}
where $\tilde\psi =\psi^\dagger\gamma_0$.

Now the vectors $p^\pm$ are null or optical because of Proposition
2, since the Weyl spinors $\varphi_\pm$ are simple or pure. The
corresponding Cartan-Weyl equations will be:
\begin{equation}
p_\mu\gamma^\mu (1\pm\gamma_5)\psi = 0
\label{Budinich:equation24}
\end{equation}
which may be expressed in Minkowski space-time if $p_\mu$ are
interpreted as generators of Poincar\'e translations: $p_\mu\to
i\frac{\partial}{\partial x_\mu}$. They identify, after
introduction of the Planck's constant, with the known wave
equation of motion for massless neutrinos.

Observe that in this unique derivation, obtained by merely
imposing the reality of the $p_\mu$ components, Minkowski
signature derives from quaternions, as may be seen already from
eqs.(\ref{Budinich:equation19}) and (19$'$) and from their correlation with
Clifford algebras, in fact notoriously $\Cl(1,3)=H(2)$ where $H$ stands for
quaternions. One might then affirm that Minkowski signature is the
image in nature of quaternions.

It is interesting to observe that if we define the electromagnetic
tensors $F$ with components
\begin{equation}
F^{(\pm )}_{\mu\nu} =\tilde\psi [\gamma_\mu ,\gamma_\nu ] (1\pm
\gamma_5)\psi
\label{Budinich:equation25}
\end{equation}
we obtain, from Cartan-Weyl eq.(\ref{Budinich:equation24}) the
Maxwell's equations in empty space\footnote{Also the inhomogeneous
Maxwell's equations in the presence of external electromagnetic
sources may be obtained from spinor
geometry~\cite{Budinich:Hestenes87}.}~\cite{Budinich:Budinich02}:
\begin{equation}
p_\mu F^{\mu\nu}_+ =0;\quad \epsilon_{\lambda\rho\mu\nu} p^\rho
F^{\mu\nu}_- =0 .
\label{Budinich:equation26}
\end{equation}

It is easy to see that also the electromagnetic potential may be
bilinearly expressed in terms of Weyl spinors and then, in the
quantised theory, the photon will result bilinear in neutrinos.
However, now the bound-state assumption will not be necessary, and
furthermore, as well known, it would not work.

In the last part of this chapter we naturally operated the
transition from $n=1$ to $n=2$, we have now only to generalize it.
To this end it is easy to show that the same results may be
obtained starting from the neutral Clifford algebra $\Cl(1,1)$
which also brings to $\Cl(1,3)$ as above.

\section{The rule: from $n$ to $n+1$}

Let $\psi_D\in S$ for $\Cl(1,2n-1)=\End S$ with generators $\g_a$.
Define the Weyl spinors $\varphi^\pm_W$ as in
eq.(\ref{Budinich:equation2}) and let them be simple or pure, then
because of Proposition 2:
\begin{equation}
p^\pm_a=\tilde\psi_D\g_a(1\pm \g_{2n+1})\psi_D,\quad a=1,2,\dots 2n
\label{Budinich:equation27}
\end{equation}
define null vectors in $\R^{1,2n-1}$.

Now we have $\varphi^+_W\oplus \varphi^-_W=\psi_D$  and
correspondingly: $p^+_a+p^-_a=p_a = \tilde\psi_D\g_a\psi_D$ which
are the components of a non null vector, which however is the
projection in $\R^{1,2n-1}$   of a null vector of $\R^{1,2n+1}$
with real components:
\begin{equation}
P_A=\tilde\Psi\G_A(1+\G_{2n+3})\Psi,\quad A=1,2,\dots 2n+2
\label{Budinich:equation28}
\end{equation}
with $\Psi\in S$, $\Cl (1, 2n + 1) =\End S$, generated by $\G_A$, which
defines the Weyl  (simple) spinors.
\begin{equation}
\psi^\pm_W=\frac{1}{2}(1\pm\G_{2n+1})\Psi_D
\label{Budinich:equation29}
\end{equation}

It may be shown~\cite{Budinich:Budinich02} that the real
components $P_A$ may be written in the form:
$P_a=p_a=\tilde\psi_D\g_a\psi_D;
P_{2n+1}=i\tilde\psi_D\g_{2n+1}\psi_D;
P_{2n+2}=\tilde\psi_D\psi_D$.

Therefore the rule is:
\begin{equation}
\varphi^+_W\oplus\varphi^-_W=\psi_D
\label{Budinich:equation30}
\end{equation}
and
\begin{equation}
p^+_a\oplus p^-_a\hookrightarrow P_A=\{ p_a,P_{2n+1},P_{2n+2}\} .
\label{Budinich:equation31}
\end{equation}

The corresponding Cartan-Weyl equation is:
\begin{equation}
P_A\G^A(1\pm\G_{2n+3})\Psi_D=0,\quad A=1,2,\dots 2n+2
\label{Budinich:equation32}
\end{equation}
which, because of Proposition 3, may be set in the
form~\cite{Budinich:Budinich02}:
\begin{equation}
(P^a\g_a^{(m)}+iP_{2n+1}\g_{2n+1}^{(m)}\pm P_{2n+2})\psi^{(m)}=0,\ \
m=0,1,2,3
\label{Budinich:equation33}
\end{equation}
and a similar one for the signature $(2n+1,1)$:
$$
(P^a\g^{(m)}_a+P_{2n+1}\g^{(m)}_{2n+1}\pm
iP_{2n+2})\psi^{(m)}=0,\quad m=0,1,2,3\hspace{3.5cm} \eqno(33')$$

To interpret physically eqs.(\ref{Budinich:equation33}) and
(33$'$) we have only to interpret the first four $P_\mu$ as
generators of Poincar\'e translations: $i\frac{\partial}{\partial
x_\mu}$ by which Minkowski space-time is generated as a
homogeneous space, and then consider both the spinor $\psi^{(m)}$
and $P_j$, with $j>5$ as functions of $x_\mu$; the latter
representing external fields (bilinear in spinors).

We see then that the rule which derives from the generalization of
the natural first step of chapter 3, and coherent with Cartan's
conjecture, foresees steadily the appearance of lorentzian
signatures for all values of $n$.

\begin{remark} Observe
that in eqs.(\ref{Budinich:equation33}) or (33$'$) the $P_A$
components, have to define a null vector of $\R^{1,2n+1}$ or
$\R^{2n+1,1}$, in order to have for $\psi^{(m)}$ non null
solutions. This condition is geometrically implied by Proposition
2 if we adopt for our momentum space Cartan's conjecture on the
fundamental role of simple spinors. The corresponding momentum
space will then result compact and equivalent to the Poincar\'e
invariant mass-sphere:
\begin{equation}
\pm P_\mu P^\mu =M^2_n=P^2_5+P^2_6+\dots P^2_{2n+2}
\label{Budinich:equation34}
\end{equation}
whose radius increases with $n$; that is, with the dimension
$2^{n-2}$ of the fermion  multiplet we are dealing with. In such
space, in the quantized field theory there will not be
ultraviolet divergences.
\end{remark}

Observe that in chapter 3 we naturally operated the transition
from  $n=1$ to $n=2$ which is a particular case: for $n=1$, of the
general rule defined in this chapter. In that case in terms of
Clifford algebras we went, de facto, from $\Cl(1,1)$ to
$\Cl(1,3)$.

Now let us recall the Bott periodicity theorem on Clifford
algebras, stating that: $\Cl(n+8,m)=\Cl(n,m+8)=\Cl(n,m)\otimes\R
(16)$.

If we apply it to our case we should increase $n$, step by step
from $n=1$ up to $n=5$ arriving at
\begin{equation}
\Cl(1,1)\otimes \R (16) =\Cl(1,9)=\Cl(9,1)=\R (32)
\label{Budinich:equation35}
\end{equation}
since $\Cl(1,1)= \R (2)$, after which the cycle, because of the periodicity
theorem, will be repeated. Now it happens that it is
precisely $\R^{1,9}$ the higher dimensional space which is generally
adopted to explain the main features of elementary particle physics. This
coincidence might not be accidental.

We will now concisely list the features which are naturally and
uniquely emerging at the various steps of our construction and,
somehow surprisingly, we will see that they reproduce most of the
known elementary particle properties. Several of these appear as
due to the known correlations of Clifford algebras with division
algebras.

\section{The steps from $n=2$ to $n=5$}

We will now study eqs.(\ref{Budinich:equation33}) or (33$'$) for
increasing $n$: from $2$ to $5$, and try to interpret them as
equations of motions (for $p_\mu\to i\frac{\partial}{\partial
x_\mu}$) for fermions or fermions multiplets. In this journey we
will naturally find the equations, traditionally postulated ad
hoc, in order to represent elementary phenomena of fermions. We
will find them more or less in the same order as they were
historically postulated: from the isospin symmetry for nuclear
forces up to the $SU(3)$ ones for quarks; however now they derive
from pure spinor geometry and, in particular, internal symmetries
appear to originate from the division algebras correlated with the
corresponding Clifford algebras, but we will also obtain
indications on the possible geometrical
origin of charges, families and some other features.\\

\noindent{$\mathbf{n=2}$}

\noindent Eq.(33$'$) for $\Cl(3,1)=R(4)$ may represent Majorana
spinors~\cite{Budinich:Budinich02}.\\

\noindent{$\mathbf{n=3}$}

\noindent Eq.(33$'$) for $m=0$ is: $(p_a\G^a+p_7\G_7+ip_8)$ $N=0$
where $\G_a(a=1,2\dots 6)$ are the generators of $\Cl(5,1)$, $N=
\begin{pmatrix} \psi_1\\ \psi_2\end{pmatrix}$, since $m=0$,
is a doublet of $\Cl(3,1)$
Dirac spinors. It may be written
in the form:
\begin{equation}
(p_\mu 1\otimes\gamma^\mu +\vec\pi\cdot\vec\sigma\otimes\gamma_5 +
M)N=0 \label{Budinich:equation36}
\end{equation}
where the pion field $\vec\pi =\frac{1}{8}\tilde
N\vec\sigma\otimes\g_5N$ is bilinear in $N$.
Eq.(\ref{Budinich:equation36}) well represents (for $p_\mu\to
i\frac{\partial}{\partial x_\mu}$) the pion nucleon equation of
motion with isospin symmetry $SU(2)$ clearly of quaternionic
origin. However now $SU(2)$ is not the covering of $SU(3)$; it is
generated by the reflection operators $\G_5,\G_6,\G_7$, which are
the same as those of the conformal
group~\cite{Budinich:Budinich02}. Obviously one might also obtain from
eq.(\ref{Budinich:equation36}) the equation of motion for the pion
field in a similar way as Maxwell's equations were derived from
massless neutrino eq.(\ref{Budinich:equation24}) in Chapter 3.

>From eq.(\ref{Budinich:equation36}) one may also derive the role
of complex numbers at the origin of the electric charge. In fact,
let us write (\ref{Budinich:equation36}) explicitly for $\psi_1$
and $\psi_2$:
\begin{gather}
     \left( p_\mu \gamma^\mu + p_7 \gamma_5 + ip_8 \right) \psi_1 +
     \gamma_5 \left( p_5 - i p_6 \right) \psi_2 = 0\, ,  \nonumber\\
     \left( p_\mu \gamma^\mu - p_7 \gamma_5 + ip_8 \right) \psi_2 +
     \gamma_5 \left( p_5 + i p_6 \right) \psi_1 = 0\, .
\label{Budinich:equation37}
\end{gather}
after defining: $p_5\pm ip_6=\rho e^{\pm \frac{i\omega}{2}}$,
where $\rho^2=p^2_5+p^2_6$, we obtain:
$$\begin{array}{rl}
  \left( p_\mu \gamma^\mu + p_7 \gamma_5 + ip_8 \right) e^{ i
  \frac{\omega}{2}} \psi_1 + \gamma_5 \rho \psi_2 &= 0\\
\left(
  p_\mu \gamma^\mu - p_7 \gamma_5 + ip_8 \right) \psi_2 + \gamma_5 \rho
  e^{i \frac{\omega}{2}} \psi_1 &= 0\hspace{6.5cm}
\end{array}
\eqno(37')
$$
which presents a $U(1)$ phase invariance of $\psi_1$ generated by
$J_{56}=\frac{1}{2}[\G_5,\G_6]$ which being local induces a
covariant derivative and (\ref{Budinich:equation36}) may be written in
the form:
\begin{equation}
  \left\{ \gamma_\mu \left[ i \frac{\partial}{\partial x_\mu} +
  \frac{e}{2} \left( 1 -i \G_5 \G_6 \right) A_\mu \right] + \vec{\pi}
  \cdot \vec{\sigma} \otimes \gamma_5 + M \right\} \begin{pmatrix}
  p\cr n \ \end{pmatrix} = 0
  \label{Budinich:equation38}
\end{equation}
where we set $\psi_1=p, \psi_2=n$, well representing the
proton-neutron doublet interacting with the pion and with the
electromagnetic potential $A_\mu$.\\

\noindent{$\mathbf{n=4 }$}

\noindent Eq.(33$'$) for $m = 0$ is:
\begin{equation}
(p_AG^A+p_9G_9+ip_{10})\Theta =0,\quad A=1,2,\dots 8
\label{Budinich:equation39}
\end{equation}
where $G_A$ are generators of $\Cl(7,1)$ and $\Theta
=\begin{pmatrix} N_1\\ N_2\end{pmatrix}$ is a doublet of
$\Cl(5,1)$ Dirac spinors.

It is easy to see~\cite{Budinich:Budinich02} that $N_1$ (or $N_2$)
presents a $U(1)$ covariance generated by
$J_{78}=\frac{1}{2}[G_7,G_8]$ at the origin of a charge for $N_1$
from which $N_2$ (or $N_1$) is free. If interpreted as the strong
charge, the $N_1$ may represent a baryon doublet and $N_2$ a
lepton one (from which the similarity of lepton-baryon families,
discussed below).

But we may also obtain a non abelian gauge field. We will
illustrate it in the case of the possible geometrical origin of
the electroweak model. In fact supposing $\Theta =\begin{pmatrix}
L_1\\ L_2\end{pmatrix}$ to represent leptons (see next step:
$n=5$) and, starting from eq.(\ref{Budinich:equation33}), assume
the generators $G_A$ of the form (we selected the signature
($1,7)$):
\begin{gather}
G_\mu =1_2\otimes\sigma_3\otimes\g_\mu ; G_{5,6}=i\sigma_1\otimes
\sigma_{1,2}\otimes 1_4;\nonumber \\
G_7=i\sigma_1\otimes\sigma_3\otimes\g_5;
G_{8,9}=i\sigma_{2,3}\otimes 1_2\otimes\g_5 \label{Budinich:equation40}
\end{gather}
Then from eq.(\ref{Budinich:equation39}) we may easily derive:
\begin{gather}
p_\mu\sigma_3\otimes\g^\mu L_{1R}+(p_9-p_{10})L_{1L}+
(p_8+\frac{i}{2}\boldsymbol{\omega}\cdot\boldsymbol{\sigma})
L_{2L}=0\nonumber \\
p_\mu\sigma_3\otimes\g^\mu L_{2R}+(p_9+p_{10})L_{2L}+
(p_8-\frac{i}{2}\boldsymbol{\omega}\cdot\boldsymbol{\sigma})
L_{1L} \label{Budinich:equation41}
\end{gather}
where $\frac{1}{2} \boldsymbol{\omega}\cdot\boldsymbol{\sigma}
=p_5\sigma_1+p_6\sigma_2+p_7\sigma_3$ and $L_{jL},L_{jR}$;
represent, for $j=1,2$, left-handed and right-handed projections
of $L_1$ and $L_2$ respectively. Now defining
\begin{equation}
p_8\pm\frac{i}{2} \boldsymbol{\omega}\cdot\boldsymbol{\sigma}
=\rho e^{\pm\frac{i}{2}
\boldsymbol{\omega}\cdot\boldsymbol{\sigma}} =\rho
e^{\pm\frac{q}{2}} \label{Budinich:equation42} \end{equation}
where $q=i\boldsymbol{\omega}\cdot\boldsymbol{\sigma}$ represents
an imaginary quaternion, it is easy to see that $L_{1L}$ presents
a phase invariance $e^{i\frac{q}{2}}\to e^{i\frac{q'}{2}}$ from
which $L_{2L}$ is free. Since $q$ is local this gives rise to a
non abelian covariant derivative.
\begin{equation}
D_\mu =\frac{\partial}{\partial
x_\mu}+i\boldsymbol{\sigma}\cdot\mathbf{w_\mu} \label{Budinich:equation43}
\end{equation}
which if, applied to the electron $e$ and left-handed neutrino
$\nu_L$ doublet: $L_1=\begin{pmatrix}e\\ \nu_L\end{pmatrix}$ gives
origin to the equation:
\begin{equation}
\begin{pmatrix} i\frac{\partial}{\partial x_\mu}+m\end{pmatrix}
\begin{pmatrix} e\\ \nu_L\end{pmatrix} +\boldsymbol{\sigma}\cdot\mathbf{w_\mu}
\g^\mu \begin{pmatrix} e_L\\ \nu_L\end{pmatrix} +B_\mu\g^\mu
e_R=0, \label{Budinich:equation44} \end{equation} which is the
strating equation for the electroweak model
(eq.(\ref{Budinich:equation44}) may
also be obtain directly from Proposition 3~\cite{Budinich:Budinich02}).\\

\noindent{$\mathbf{n=5}$}

\noindent Eq.(33$'$) for $m=0$ is:
\begin{equation}
(p_\alpha \cg^\alpha + P_{11}\cg_{11} +iP_{12})\Phi =0, \quad
\alpha = 1,2,\dots 10
 \label{Budinich:equation45} \end{equation}
With $\cg_\alpha$ generators of
$\Cl(9,1)$ and $\Phi=\begin{pmatrix}\Theta_1\\
\Theta_2\end{pmatrix}$ a doublet of $\Cl(7,1)$ Dirac spinors.
Again, $\Theta_1$ has a $U(1)$ phase invariance generated by
$J_{9,10}=\frac{1}{2}[\cg_9,\cg_{10}]$ not presented by
$\Theta_2$; we will then assume $\Theta_1:=\Theta_B$ to represent
a quadruplet of baryons and $\Theta_2:=\Theta_{{\cal L}}$  a
quadruplet of leptons.

\section{The Baryon quadruplet $\Theta_B$}

$\Theta_B$ obey eq.(\ref{Budinich:equation39}), therefore it
defines the invariant mass of eq.(\ref{Budinich:equation34}):
\begin{equation}
-p_\mu p^\mu =M^2_4=p^2_5+p^2_6+p^2_7+p^2_8+p^2_9+p^2_{10}
\label{Budinich:equation46}
\end{equation}
which defines a sphere $S_5$ presenting a symmetry $SO(6)$
orthogonal to the Poincar\'e group. Therefore the maximal internal
symmetry for the baryon quadruplet $\Theta_B$ could be $SU(4)$.
However before this there is an internal symmetry originating from
the maximal complex division algebra: that of octonions. In fact
it was shown~\cite{Budinich:Sudbery87} that $\Spin (1,9) = \Spin
(9,1)\cong SL(2,\mathbf{o})$ where $\mathbf{o}$ stands for
octonion (even if the isomorphism is restricted to the
infinitesimal groups).

In order to set it in evidence let us first remind that
eq.(\ref{Budinich:equation39}) for $\Theta_B$ derives from the
Cartan-Weyl eq.(\ref{Budinich:equation32}) which in our case is:
\begin{equation}
p_\alpha\cg^\alpha (1+\cg_{11})\Phi =0 ,\quad \alpha = 1,2,\dots
10 \label{Budinich:equation47}
\end{equation}
and, since $\Theta_B$ presents an $U(1)$ local covariance it will
determine a covariant derivative in space-times such that
eq.(\ref{Budinich:equation47}) may be written in the form:
\begin{equation}
\left[ i\left(\frac{\partial}{\partial x_\mu}
-igA^\mu_{(m)}\right)\cg^{(m)}_\mu + \sum^9_{j=5}
p_j\cg^{(m)}_j+ip_{10}\right]\Phi^{(m)} =0
\label{Budinich:equation48}
\end{equation}
where the indices $m$ derive from Proposition 3 as in eq.(33$'$).

It may be shown~\cite{Budinich:Budinich02} that the isomorphism
with the octonion algebra may be represented through the
$\cg_\alpha$ matrices. Precisely $\cg^{(n)}_\mu$ and $\cg_{6+n}$
for $n=1,2,3$ may represent the first 3 octonion imaginary unit
$e_1,e_2,e_3$ respectively while $i\cg_{11}$ represent the seventh
one $e_7$, such that the projector $\frac{1}{2} (1+\cg_{11})$
selects a particular direction in octonion space. This then
reduces the automorphism group $G_2$ of octonions to $SU(3)$. In
this way the so-called complex octonions may be defined:
\begin{equation}
U_\pm = (1\pm\cg_{11}); \ \ V^{(n)}_{\mu\pm} =\cg^{(n)}_\mu U_\pm
; \ \ V^{(n)}_\pm =\cg_{6+n}U_\pm . \label{Budinich:equation49}
\end{equation}

They represent an $SU(3)$ invariant algebra~\cite{Budinich:Gursey&Tze96}.
Precisely $U_\pm$ transform as
singlets and $V^{(n)}_{\mu +}$ and $V^{(n)}_+$ transform as the
$(3)$ representation of $SU(3)$ while $V^{(n)}_{\mu -}$ and
$V^{(n)}_-$ as the $(\bar 3)$ one. In this way
eq.(\ref{Budinich:equation48}) may be set in the form:
\begin{equation}
\left[ i\left(\frac{\partial}{\partial x_\mu}
-igA^\mu_{(n)}\right) V^{(n)}_{\mu +}+
p_5\cg_5+p_6\cg_6+\sum^3_{n=} p_{6+n} V^{(n)}_+ip_{10}\right]
U_+\Phi =0 , \label{Budinich:equation50}
\end{equation}
where the term with $V^{(n)}_{\mu +}$ may be interpreted to
represent $SU(3)$ color and the one with $V^{(n)}_+$ represents
$SU(3)$ flavours. Eq.(\ref{Budinich:equation50}) has no direct
physical interpretation since the vector component like $p_{6+n}$
are bilinear in spinors. Formally we could obtain $SU(3)$ flavour
covariance if we would define $p_{6+n} =\Phi^\dagger\cg_0V^{(n)}_-\Phi$.

But it should also be possible to obtain the known Gell-Mann
$3\times 3$ representation of the pseudo-octonion
algebra~\cite{Budinich:Okubo95} by acting with the 3 operators
corresponding to $V^{(n)}_+$ on Cartan-standard spinors, or
equivalently, on the vacuum of a Fock representation of spinor
space as in Ref.5 to obtain, as minimal left ideals, 3 spinors
representing quarks, as will be discussed elsewhere. In this way
the $SU(3)$ symmetry both of flavour and color might be obtained
in the framework of the algebraic theory of spinors, as already
obtained by other authors~\cite{Budinich:Dixon94},
\cite{Budinich:Trayling&Baylis01}.

According to the present geometrical scheme a fourth quark should
exist presenting with the other 3 on $SU(4)$ symmetry. It could be
discovered at higher energies.

\section{The lepton quadruplet $\Theta_{{\cal L}}$}

We have seen that $n\to n+1$ means doubling the dimension of
spinor space and adding two more terms to the equations of motion.
This means that in our geometrical scheme dimensional reduction
means $n\to n-1$ and is equivalent to reducing to one half the
dimension of spinor space and decoupling of the equations of
motions. Therefore the quadruplet $\Theta_{{\cal L}}$, missing
strong charge, will have to be reduced to a doublet; as in fact it
appears in nature where lepton always appear as charged-neutral
doublets. Now it may be easily seen that this dimensional
reduction, because of Proposition 3, may be operated in 3 non
equivalent ways giving origin to 3 lepton-neutrino families
differing in the values of the invariant
masses~\cite{Budinich:Budinich02}. And in nature electron, muon
and $\tau$ lepton seem to differ mainly in masses. It may be shown
that the 3 families may be correlated with the 3 imaginary units
of quaternions~\cite{Budinich:Dray&Manogue99}.

Observe that, if $\Theta_{{\cal L}} = \begin{pmatrix} L_1\\
L_2\end{pmatrix}$ where $L_1$ and $L_2$ are doublets of leptons,
as we have seen in Chapter 4, for $n=4$, it is foreseen that if
$L_1$ presents electroweak interactions, $L_2$ do not. Therefore
$L_2$ needs a further dimensional reduction and decoupling of the
equations of motion; that is a reduction from $n=3$ to $n=2$ and
$n=1$ obtaining the equations for Majorana fermions and neutrinos.
They could be the candidates for the explanation of the origin of
black matter~\cite{Budinich:Budinich02}.

\section{Concluding remarks and outlook}

We have seen how Cartan's simple spinors are appropriate to
represent most of the observed properties of fermions. They
plainly explain internal symmetries as due to their correlation
with Clifford algebras; they explain the origin of charges and the
tendency of fermions to appear in charged-neutral pairs or pairs
of multiplets, as the baryon-leptons ones, with similar
properties; in particular the origin of families.

In this framework the main role of simplicity is to render compact
(due to Proposition 2) the momentum spaces where the equations of
motion are naturally formulated. This implies that the
corresponding quantized theory (in second quantization) will be
free from one of the most severe difficulties of quantum field
theories; that of ultraviolet divergences.

In this preliminary approach there appear several aspects which
deserve further study. One is the physical meaning of
eq.(\ref{Budinich:equation34}) defining invariant masses (which
implicitly contains also the charges) and the possibility it may
offer to compute their values. Another is the constraint relations
which simple or pure spinors must obey. In our case they are in
number $66$ for the spinor appearing for $n=5$ in Chapter 5 (and
in our interpretation representing 4 baryons and 4 leptons), and
in number $10$ for the $\Cl(1,9)$-pure spinors considered in
Chapter 6 (in our interpretation representing 4 baryons). It is
interesting to observe that this case has already been studied
with unexpected positive results in superstring
theory~\cite{Budinich:Matone}.

However, this approach might present further interesting aspects.
In fact let us assume as a postulate what it indicates: that for
the appropriate description of the wave mechanics of fermions we
need the geometry of Cartan's simple spinors. Now this mechanics
is generally conceived as the constituent of classical mechanics
of macroscopic bodies of which fermions are in turn the elementary
constituents. Let us now remind that, classical mechanics is well
representable with euclidean geometry (or with its Riemanian
generalizations) in space-time and, in this framework, the
physical interpretation of quantum mechanics presents known and
widely discussed difficulties. But now there could be another
possibility, once again obtained by shifting the attention from
physics to geometry. In fact if the geometry to be adopted for
dealing with fermions is that of Cartan's simple spinors, then,
following Cartan, those spinors might represent the elementary
constituents of euclidean geometry and the problem may be shifted
on how, from simple spinors, one may construct the elements of
euclidean geometry. All this may be synthetically represented in
the table above: where with Perceptible World we mean that which can be directly
perceived with out senses and is then accessible to our ordinary
intuition. The arrows indicate embeddings.

\begin{table}
\begin{center}
\begin{tabular}{l|lll}
&PHYSICAL&MECHANICS&GEOMETRICAL\\
&OBJECTS&&INSTRUMENTS\\ \hline &&&\\
PERCEPTIBLE&Macroscopic&Classical&Euclidean\\
WORLD&bodies&mechanics&geometry\\
&&&\\
&$\uparrow$&$\uparrow$&$\uparrow$\\
&1&2&3\\
&&&\\
ELEMENTARY&Fermions&Quantum&Cartan's\\
CONSTITUENTS&&mechanics&simple spinors\\ \hline
\end{tabular}
\end{center}
\end{table}

For the embedding 1 there is no problem, (basons are bilinears of
fermions). The embedding 2 is problematic; it could be substituted
with the embedding 3 following Cartan's conjecture\footnote{This,
among others, would set in evidence a striking parallelism between
physics and geometry.}.

In other worlds the problem of how quantum mechanics may be
embedded in classical mechanics, and then understood in its frame,
would be shifted from physics to geometry.

Now notoriously, this problem when dealt with in the frame of the
embedding 2, that is in the frame of physics, gives rise to known
paradoxes; that is to propositions contradicting our ordinary
intuition. If we shift the problem to the embedding 3 that is to
how one may construct euclidean geometry from that of Cartan's
simple spinors we will have to deal only with abstract
mathematical and geometrical objects for which we do not need the
steady control of our ordinary intuition\footnote{As an example,
mentioned above, a simple spinor associated with a complex
euclidean space of dimension $2n$ is equivalent (up to a sign) to
a totally null plane of dimension $n$ whose vectors are all null
and mutually orthogonal, which is mathematically correct but no
easily accessible to our ordinary intuition.}.

Furthermore, when the progress of theoretical physics is guided by
geometry or mathematics it may well happen that apparent paradoxes
may arise, as it happened in relativity with the apparent
paradoxes of dilatation of proper time\footnote{Like the one: it
takes about 3 years for light to arrive here from $\alpha$
centauri. I could arrive there tomorrow if I could dispose of a
vehicle travelling with a velocity slightly \underline{smaller}
than that of light.}, and then they are accepted as a appropriate
corrections to the errors and limitations of our ordinary
intuition.

Some tentative consequence of such a geometrization of the problem
may be set in evidence already at this preliminary stage. In fact
we have seen that with simple or pure spinors we may only obtain,
bilinearly (according to Proposition 2) euclidean null vectors;
the ordinary vectors may then be obtained only as sums or
integrals of them; and the latter may represent
strings~\cite{Budinich:Budinich&Rigoli}. This might explain the
motivation of the necessity of strings, in quantum mechanics. In
fact coherently with this approach we cannot introduce in our
quantum mechanics the concept of point-event which is a concept of
euclidean geometry in space-time valid only for classical
mechanics. Neither we can construct, via Fourier transforms, this
concept, since for that we would need an infinite momentum space
of which we do not dispose, as seen above. Therefore the resulting
quantum mechanics in space-time will have to be fundamentally non
local. The rigorous way to represent this non locality should be
studied in dealing with the geometrical problem of embedding
Cartan's simple spinors in euclidean geometry and will be dealt
with elsewhere.

\end{document}